\documentclass[final,5p]{elsarticle}

\usepackage{lineno,hyperref}
\modulolinenumbers[5]

\journal{Physics Letters A}

\let\today\relax
\makeatletter
\def\ps@pprintTitle{%
    \let\@oddhead\@empty
    \let\@evenhead\@empty
    \def\@oddfoot{\footnotesize\itshape
         {Submitted preprint} \hfill\today}%
    \let\@evenfoot\@oddfoot
    }
\makeatother

\usepackage{graphicx}
\usepackage{bm,bbm}
\usepackage[dvipsnames]{xcolor}

\usepackage{amsmath}
\usepackage{amsfonts}
\usepackage{amssymb}
\usepackage{epstopdf}

\newcommand{\One}{\mathbbm{1}}
\newcommand{\rmd}{\mathrm{d}}
\newcommand{\rme}{\mathrm{e}}
\newcommand{\rmi}{\mathrm{i}}

\newcommand{\la}{\langle}
\newcommand{\ra}{\rangle}

\newcommand{\ie}{i.e.}
\newcommand{\ket}[1]{|#1\rangle}
\newcommand{\bra}[1]{\langle #1|}
\newcommand{\proj}[2]{{\vert #1 \rangle \langle #2 \vert}}

\begin{document}

\begin{frontmatter}

\title{Positivity and complete positivity of differentiable quantum processes}

\author[GDLaddress]{Gustavo Montes Cabrera}
%\address{Departamento de F\'\i sica, Universidad de Guadalajara, Guadalajara, 
%   Jal\'\i sco, M\' exico.}

\author[UNAMaddress]{David Davalos}

\author[GDLaddress]{Thomas Gorin\corref{mycorrespondingauthor}}
\cortext[mycorrespondingauthor]{Corresponding author}
\ead{thomas.gorin@cucei.udg.mx}

\address[GDLaddress]{Departamento de F\'\i sica, Universidad de Guadalajara, 
   Guadalajara, Jal\'\i sco, M\' exico.}
\address[UNAMaddress]{Instituto de F\'\i sica, Universidad Nacional 
   Aut\' onoma de M\' exico, Ciudad de M\' exico, M\' exico.}

\begin{abstract}
We study quantum processes, as one parameter families of differentiable
completely positive and trace preserving (CPTP) maps. Using different 
representations of the generator, and the Sylvester criterion for positive 
semi-definite matrices, we obtain conditions for the divisibility of the 
process into completely positive (CP-divisibility) and positive 
(P-divisibility) infinitesimal maps. Both concepts are directly related 
to the definition of quantum non-Markovianity. For the single qubit case we 
show that CP- and P-divisibility only depend on the dissipation matrix in the 
master equation form of the generator. We then discuss three classes of 
processes where the criteria for the different types of divisibility result in 
simple geometric inequalities, among these the class of non-unital anisotropic 
Pauli channels.
\end{abstract}

\begin{keyword}
Quantum process \sep Divisibility \sep Quantum non-Markovianity

%\texttt{elsarticle.cls}\sep \LaTeX\sep Elsevier \sep template
%\MSC[2010] 00-01\sep  99-00
\end{keyword}

\end{frontmatter}

%\linenumbers

\section{\label{I} Introduction}

Non-Markovianity of quantum processes has been a topic of increasing interest
during approximately the last ten years~\cite{Breuer12,RHP14,LiHaWi18}. 
Starting with the papers by Breuer et al.~\cite{BrLaPi09} and Rivas et 
al.~\cite{RiHuPl10}, a definition of quantum Markovianity has been reduced to 
the question whether all intermediate quantum maps are physically realizable; 
this induces a characterization that is more closely related to the 
Chapman-Kolmogorov condition than to the full definition of classical 
Markovianity~\cite{BrePet02}. For differentiable quantum processes, the 
question of divisibility into physically realizable quantum maps can be further
reduced to the analysis of the time dependent generator of the process. This is
the approach taken for this work.

The concept of divisibility has been introduced in 
Refs.~\cite{WolCir08,WoEiCuCi08}. In its original form, it refers to the 
condition that all intermediate maps are completely positive (CP-divisibility). 
However, one may as well consider P-divisibility, where it is sufficient that 
the intermediate maps are positive~\cite{BeFlPi04,ChrMan14}. If an intermediate 
map is positive but not completely positive, one may observe information 
backflow for entangled states between system and some ancillary system, but not 
in the system alone~\cite{RiHuPl10,Chr11}

% Our objective
In this work, we derive general criteria for positivity and complete 
positivity. In particular, for single qubit processes we show that both, 
CP-divisibility and P-divisibility conditions, only depend on the dissipation 
matrix of the master equation. We identify three different classes of single 
qubit processes, where the criteria for CP- and P-divisibility are reduced to 
simple explicit geometric inequalities. One of these classes consists of 
processes where the Choi-matrix has the shape of an $X$ (it means that all 
non-zero elements are located on the diagonal or the anti-diagonal). Many 
examples considered in the literature of quantum non-Markovianity are of this 
type. A second class consists of those processes, where the Choi-matrix has the 
shape of an $O$. The third class is that of the non-unital anisotropic Pauli 
channels. While criteria applicable to the generators have been studied in the 
context of CP-divisibility, see for instance Ref.~\cite{Hall08}, this has 
rarely been done for P-divisibility. 

Explicit analytical criteria are valuable for the construction of Markovian 
approximations to a non-Markovian process as proposed in 
Ref.~\cite{WoEiCuCi08} and more specifically in Ref.~\cite{RiHuPl10}. Another 
area of applications is that of quantum process 
tomography~\cite{NieChu00, Bou03, Dom17}, where it is important to identify 
the independent parameters which are to be determined. Finally, it may be of 
interest to identify quantum channels, which are P-divisible but not 
CP-divisible as processes where non-Markovianity may be interpreted as a 
genuine quantum effect~\cite{ChruPas17}. 

Our work relies on a few general results which have been derived previously.
The most important ones are (i) the Kossakowski theorem, which establishes the
equivalence between positivity and contractivity (for the domain of Helstrom
matrices)~\cite{Chr11} (and references therein); (ii) necessary and sufficient
criteria which can be applied directly to the time dependent generator of
the quantum process, Ref.~\cite{BeFlPi04} for positivity and
Ref.~\cite{WoEiCuCi08} for complete positivity; and finally (iii) Sylvester's 
criterion for definite and semi-definite positivity~\cite{Meyer00,Pru86}.

The paper is organized as follows: In Sec.~\ref{C} we discuss the description 
of quantum processes in terms of their generators and the general conditions 
for CP- and P-divisibility in terms of the generator. In Sec.~\ref{S} we 
analyze these conditions for general single qubit processes. In Sec.~\ref{E} we 
present classes of single qubit processes with analytically solvable conditions 
for P- and CP- divisibility. In Sec.~\ref{F} we present our conclusions.

\section{\label{C} Differentiable quantum processes}

In this section we introduce differentiable quantum processes and the 
definitions of P-divisibility or CP-divisibility. For both types of
divisibility, we present criteria which can be applied directly to the 
generator of the quantum process in question.

\subsection{\label{CP} Processes and generators}

Let us denote a quantum process $\Lambda_t$, ($\forall t\in\mathbbm{R}_0^+$),
as a one-parameter family of differentiable (with respect to $t$) completely 
positive and trace preserving linear maps (CPTP-maps), with $\Lambda_0 = \One$, 
the identity. For simplicity, we assume that the corresponding Hilbert space is 
of finite dimension, $\dim(\mathcal{H}) = d < \infty$. The quantum process
$\Lambda_t$ can be defined equivalently by the generator $\mathcal{L}_t$, such 
that
\begin{align}
\frac{\rmd}{\rmd t}\, \Lambda_t = \mathcal{L}_t\; \Lambda_t \; , \qquad
\Lambda_0 = \One \; .
\end{align}
One natural question to ask would be the following: What are the properties to 
be fulfilled by $\mathcal{L}_t$ in order to produce a valid quantum process of
CPTP maps (very recently this question has been addressed in 
Ref.~\cite{Kol18}). In the present work, we have a different objective. 
Assuming that $\mathcal{L}_t$ generates a valid quantum process, we ask 
whether that process is CP-divisible and/or P-divisible.

Note that for a given quantum process $\Lambda_t$, we can compute its generator 
as
\begin{align}
\mathcal{L}_t =  \frac{\rmd\Lambda_t}{\rmd t}\; \Lambda_t^{-1} \; .
\label{CP:Generator}
\end{align}
In what follows we will assume that $\Lambda_t$ is invertible. It is common 
that in a given quantum process, $\Lambda_t$ is non-invertible at isolated 
points in time. If this is the case, one has to proceed with 
care~\cite{ChrRiSto18}. In order to derive P-divisibility and CP-divisibility 
criteria in terms of the generator, we need to relate it to the intermediate 
quantum map,
\begin{align}
\Lambda_{t+\delta,t} = \Lambda_{t+\delta}\; \Lambda_t^{-1} \; .
\label{CP:intermedmap}\end{align}
This can be achieved by considering an infinitesimal intermediate time step. In
that case, it holds that
\begin{align}
\mathcal{L}_t = \lim_{\delta\to 0} \delta^{-1}\, \big (\,
   \Lambda_{t+\delta,t} - \One \, \big )\; .
\label{CP:LocalInterMap}\end{align}

\paragraph{Choi-matrix representation}
A direct method to represent linear quantum maps (this includes generators such
as $\mathcal{L}_t$) consists in embedding the state space into the vector space 
$\mathcal{M}^{d\times d}$ of complex quadratic matrices of dimension $d$. In
such case, the elements $\{ \, |i\ra\la j| \, \}_{1\le i,j\le d}$ form a 
convenient orthonormal basis with respect to the Hilbert-Schmidt scalar product
$\la A, B\ra = {\rm tr}(A^\dagger\, B)$. Then, we define the Choi-matrix 
representation~\cite{Choi75} of any linear map $\Lambda$ in 
$\mathcal{M}^{d\times d}$ as
\begin{align}
C_\Lambda = \sum_{i,j} |i\ra\la j| \otimes \Lambda[\, |i\ra\la j|\, ] \; .
\label{CP:ChoiMaRep}\end{align}
In practice this matrix is a $d\times d$ matrix of block-matrices from
$\mathcal{M}^{d\times d}$ which are the images of the basis elements 
$|i\ra\la j|$ under the map $\Lambda$. The remarkable properties of this 
representations are the following: $C_\Lambda=C_\Lambda^\dagger$ iff 
$\Lambda[\Delta^\dagger]=\Lambda[\Delta]$ for every bounded operator $\Delta$, 
$C_\Lambda \geq 0$ iff $\Lambda$ is complete positive and 
$\text{tr} \left( C_\Lambda \right)=d$ if $\Lambda$ preserves the 
trace~\cite{Choi75,HeiZimBook11}.

\paragraph{Master equation}
The generator obtained in Eq.~(\ref{CP:LocalInterMap}) preserves 
Hermiticity by construction, thus we can bring it to the following standard 
form~\cite{WoEiCuCi08,Evans1977} (see~\ref{sec:canonicalform} for a detailed 
derivation):
\begin{align}
&\frac{\rmd}{\rmd t}\, \varrho = \mathcal{L}_t[\varrho]\; , 
\label{CP:LindMastEq}\\
&\mathcal{L}_t[\varrho] = -\rmi\, [H,\varrho] + \sum_{i,j=1}^{d^2-1}
   D_{ij}\, \Big ( F_i\, \varrho\, F_j^\dagger - \frac{1}{2}\, \big\{ 
   F_j^\dagger F_i\, ,\, \varrho \big\}\, \Big ) \; .
\notag\end{align}
In this expression, Planck's constant $\hbar$ has been absorbed into the 
Hamiltonian $H$. The matrix $D$ is hermitian, and the set 
$\{ F_i\}_{1\le i\le d^2}$ forms an orthonormal basis in the space of 
operators, such that ${\rm tr}(F_i^\dagger F_j) = \delta_{ij}$. In addition,
the operators are chosen such that ${\rm tr}(F_i) = 0$, except for the last 
element, which is given by $F_{d^2} = \One/ \sqrt{d}$. 

In this work, we will use the expression of Eq.~\ref{CP:LindMastEq} as one 
possible representation of the generator $\mathcal{L}_t$, at some arbitrary but 
fixed time $t$. We call this representation the ``master equation 
representation'' of the generator $\mathcal{L}_t$ and $D$ the ``dissipation 
matrix''.
Note that an intermediate quantum process 
$\Lambda_{t+\delta,t}\approx \One+\delta \mathcal{L}_t$ (with $\delta>0$), as 
defined in Eq.~\ref{CP:intermedmap}, is CPTP if and only if $D$ is positive 
semidefinite~\cite{Hall08,RHP14}. Moreover, it defines a one-parameter 
semigroup in the space of CPTP maps if the generator is time 
independent~\cite{Sud61,GoKoSu76,Lin76}.

\subsection{\label{CM} Markovianity: P-divisibility vs. CP-divisibility}

In this subsection we present the definitions for the P-divisibility and the
CP-divisibility of quantum processes. We use the term ``Markovianity'' in 
cases, where we want to refer to both types of divisibility, indistinctively.

\paragraph{CP-divisibility} A process $\Lambda_t$ is called CP-divisible 
if and only if the intermediate map $\Lambda_{t+\delta,t}$ as defined in 
Eq.~(\ref{CP:intermedmap}) is CPTP for all $t,\delta \in\mathbb{R}_0^+$.
generators which depend on time. In that case, it has been shown 
in~\cite{,RHP14} that a process constructed from 
Eq.~(\ref{CP:LindMastEq}) is CP-divisible if and only if $D\ge 0$ for all 
times.

Complete positivity of a quantum map $\Lambda$ is conveniently verified using 
the Choi matrix representation, introduced in Eq.~(\ref{CP:ChoiMaRep}).
Provided that $\Lambda$ preserves Hermiticity and the trace, it is CPTP if and 
only if the Choi matrix is positive-semidefinite~\cite{Choi75,HeiZimBook11},
\ie{} if it has only non-negative eigenvalues.

\paragraph{P-divisibility} A process $\Lambda_t$ is called P-divisible 
if and only if the intermediate map $\Lambda_{t+\delta,t}$
as defined in Eq.~(\ref{CP:intermedmap}) is PTP (positivity and trace 
preserving) for all $t,\delta \in\mathbb{R}_0^+$.

Positivity of a Hermiticity and trace preserving quantum map $\Lambda$ is more
complicated to verify. In that case, one has to show that 
$\Lambda[\varrho] \ge 0$ for all density matrices $\varrho$. In practice, it 
is sufficient to check the condition for all density matrices representing 
pure states. 

\paragraph{Local complete positivity} 
Follow Refs.~\cite{WoEiCuCi08}, and~\cite{Hall14}, let $C_\perp$ be a matrix 
representation of $C_\mathcal{L}$ in the subspace orthogonal to the Bell state
\begin{align}
|\Phi_{\rm B}\ra = \frac{1}{\sqrt{d}}\sum_i |ii\ra\; ,
\label{CM:defBell}\end{align}
where $d$ is the dimension of the Hilbert space. Then, a quantum process 
$\Lambda_t$ is locally CP at time $t$, if and only if 
\begin{align}
C_\perp \ge 0 \; .
\label{CM:localCP}\end{align}
Therefore the process $\Lambda_t$ is CP-divisible, if and only if it is locally 
CP for all $t\in \mathbb{R}_0^+$. Note that in  Ref.~\cite{Hall14}, it has been 
shown that $C_\perp$ is unitarily equivalent to the dissipation matrix $D$ 
(see~\ref{sec:canonicalform} for a detailed derivation).

\paragraph{Local positivity} A quantum process is locally positive at time $t$, 
if and only if for all orthogonal states 
$|\psi\ra ,\, |\phi\ra \in \mathcal{H}$ it holds that 
\begin{align}
\la\psi|\, \mathcal{L}_t[\, |\phi\ra\la\phi|\, ]\, \psi\ra \ge 0\; .
\label{CM:localP}\end{align}
Similar to the CP case, it holds that a quantum process $\Lambda_t$ is 
P-divisible if and only if it is locally positive for all 
$t\in\mathbb{R}_0^+$~\cite{BeFlPi04}. The equivalence between local positivity 
and P-divisibility follows from Eq.~(\ref{CP:LocalInterMap}):
\begin{align}
\la\psi|\, \Lambda_{t+\delta,t}[\, |\phi\ra\la\phi|\, ]\, |\psi\ra &\ge 0
\notag\\
\quad\Leftrightarrow\quad
\delta\; \la\psi|\, \mathcal{L}_t[\, |\phi\ra\la\phi|\, ]\, |\psi\ra +
   \la\psi|\phi\ra\, \la\phi|\psi\ra &\ge 0 \; .
\end{align}
In the limit $\delta\to 0$, this can only happen if $\psi$ and $\phi$ are
orthogonal, $\la\psi|\phi\ra = 0$. In fact, if $|\la\psi|\phi\ra|^2 > 0$, it
might very well be that 
$\la\psi|\, \mathcal{L}_t[\, |\phi\ra\la\phi|\, ]\, |\psi\ra < 0$ even if the 
process is P-divisible in the neighborhood of that point. 

To summarize, we may express both properties CP-divisibility and P-divisibility
in terms of local conditions which have to be fulfilled by the generator 
$\mathcal{L}_t$ for all times $t\in\mathbb{R}_0^+$. In what follows, we analyze 
these in more detail. To avoid overly cumbersome terminology, we denote 
generators which fulfill Eq.~(\ref{CM:localP}) and/or Eq.~(\ref{CM:localCP}) 
simply as ``positive'' and/or ``completely positive generators''.

\section{\label{S} Single qubit processes}

In the case of single qubit processes, the Bloch vector representation is yet
another method to represent quantum channels and their generators. In the 
following Sec.~\ref{SE} we discuss the following three representations: (i) the 
master equation, (ii) the Choi-matrix, and (iii) the Bloch vector 
representation and how they are related one-to-another. In Sec.~\ref{SC}, we 
derive explicit criteria for local positivity and local complete positivity in 
terms of the dissipation matrix $D$.

\subsection{\label{SE} Equivalent representations}

\paragraph{Choi matrix representation}
For our purposes, the Choi matrix representation will be the most useful. A 
CPTP-map $\Lambda$, which belongs to a quantum process, may be parametrized as
\begin{align}
C_\Lambda &= \begin{pmatrix}
      \Lambda[\, |0\ra\la 0|\, ] & \Lambda[\, |0\ra\la 1|\, ] \\
      \Lambda[\, |1\ra\la 0|\, ] & \Lambda[\, |1\ra\la 1|\, ] \end{pmatrix}
\notag\\
 &= \begin{pmatrix} 
   1-r_1 & y_1^* & x^* & 1-z_1^*\\
   y_1 & r_1 & z_2 & -x^*\\
   x & z_2^* & r_2 & y_2^* \\
   1- z_1 & -x & y_2 & 1-r_2\end{pmatrix} \; .
\label{SE:ChoiMap}\end{align}
The structure of $C_\Lambda$ is due to the fact that $\Lambda$ must preserve
Hermiticity and the trace. We have chosen the parametrization in such a way
that the parameters $r_1, r_2$, $y_1, y_2, x$, $z_1, z_2$ as functions of time
are all zero at $t=0$.

Note that any intermediate map $\Lambda_{t+\delta,t}$ is at least Hermiticity 
and trace preserving. Therefore, Eq.~(\ref{CP:LocalInterMap}) implies that the 
Choi-matrix representation of the generator $\mathcal{L}_t$ must be Hermitian, 
and in all blocks, the partial trace must be equal to zero. That leaves us with 
the following parametrization:
\begin{align}
C_\mathcal{L} = \begin{pmatrix}
   -q_1 & Y_1^* & X^* & -Z_1^*\\
   Y_1 & q_1 & Z_2 & -X^*\\
   X & Z_2^* & q_2 & Y_2^* \\
   -Z_1 & -X & Y_2 & -q_2\end{pmatrix} \; .
\label{SE:ChoiGen}\end{align}
In general, there is no simple relation between the parametrization used here,
and that of Eq.~(\ref{SE:ChoiMap}). This is because the expression for the 
generator $\mathcal{L}_t$ includes the inverse of $\Lambda_t$.

\paragraph{Master equation representation}
Note that every generator $\mathcal{L}_t$ of a Hermiticity and trace preserving 
quantum process, can be written in the form of Eq.~(\ref{CP:LindMastEq}), with 
Hermitian matrices $H$ and $D$. Therefore, we may calculate the 
Choi-representation of the generator, by inserting $\varrho= |i\ra\la j|$ into 
the RHS of Eq.~(\ref{CP:LindMastEq}), and compare the result to the general 
form in Eq.~(\ref{SE:ChoiGen}). For the calculation, we choose the following 
orthonormal operator basis $\{ F_i \}_{1\le i\le d^2}$:
\begin{align}
F_1 &= \frac{1}{\sqrt{2}}\big ( |0\ra\la 0| - |1\ra\la 1|\big )\; , &
F_2&= |0\ra\la 1|\; , \notag\\
F_3 &= |1\ra\la 0|\; , \quad\text{and}\quad\quad & F_4&= \One/\sqrt{2} \; .
\label{ONoperB}\end{align}
As a result, we obtain a linear one-to-one correspondence between the 
parameters used in the master equation representation and those, used in the 
Choi representation: 
\begin{align}
\begin{pmatrix} q_1\\ q_2\\ {\rm Re}Z_1\end{pmatrix} &= 
\begin{pmatrix}
0 & 0 & 1 \\
0 & 1 & 0 \\
1 & 1/2 & 1/2 \end{pmatrix}\;
\begin{pmatrix} D_{11}\\ D_{22}\\ D_{33}\end{pmatrix}\; , \notag\\
{\rm Im}\, Z_1 &= H_{22} - H_{11}\; , \notag\\
Z_2 &= D_{32}\; , \notag\\
\begin{pmatrix} Y_1\\ Y_2\\ X^*\end{pmatrix} &=
\begin{pmatrix} -\sqrt{2}/4 & \sqrt{18}/4 & -\rmi \\
 -\sqrt{18}/4 & \sqrt{2}/4 & \rmi \\
 \sqrt{2}/4 & \sqrt{2}/4 & \rmi \end{pmatrix}\; 
\begin{pmatrix} D_{12}\\ D_{31}\\ H_{21}\end{pmatrix}\; .
\label{SE:GenParams}\end{align}
As one might have expected, the quantity $H_{11} + H_{22}$ is irrelevant
for the representation of the generator, and may be set equal to zero without
loss of generality. Then, Eq.~(\ref{SE:GenParams}) is clearly an invertible 
linear system of equations.

\paragraph{Bloch vector representation}
Any qubit density matrix can be written in terms of the Pauli matrices and the 
identity matrix $\One$ as follows:
\begin{align}
\varrho = \frac{1}{2}\; \Big (\, v_0\; \One + \sum_{j=1}^3 v_j\; \sigma_j\,
   \Big ) \; ,
\label{SE:BlochSphereRep}\end{align}
where $v_0 = 1$ and $\vec v = (v_1, v_2, v_3)$ is a vector in $\mathbb{R}^3$
of norm $\|\vec v\| \le 1$. Any Hermiticity and trace preserving quantum map 
$\Lambda$ can then be written as an affine transformation~\cite{BenZyc06}
\begin{align}
\Lambda\;\; :\;\; \vec v \to \vec v' = R\; \vec v + \vec t\; , 
%R= ( \vec r^{(1)}, \vec r^{(2)}, \vec r^{(3)}) \; ,
\label{SE:LamBlochRep}\end{align}
where $R$ is a real not necessarily symmetric square matrix and $\vec t$ is a 
real three-dimensional vector. The coefficients of $R$ and $\vec t$ are given
by
\begin{align}
t_j = \frac{1}{2}\; {\rm tr}\big (\, \sigma_j\, \mathcal{L}_t[\, \One\, ]\,
   \big )\; , \quad
R_{jk} = \frac{1}{2}\; {\rm tr}\big (\, \sigma_j\, 
   \mathcal{L}_t[\, \sigma_k\, ]\, \big )\; .
\end{align}
For the generator with the Choi-matrix representation given in 
Eq.~(\ref{SE:ChoiGen}), we find
\begin{align}
R &= \begin{pmatrix}
   {\rm Re}(Z_2 - Z_1) & {\rm Im}(Z_1 + Z_2) & {\rm Re}(Y_1 - Y_2) \\
   {\rm Im}(Z_2 - Z_1) & -\, {\rm Re}(Z_1 + Z_2) & {\rm Im}(Y_1 - Y_2)\\
   2\, {\rm Re}(X) & -2\, {\rm Im}(X) & -q_1 -q_2\end{pmatrix} \; , \notag\\
\vec t &= \begin{pmatrix}
   {\rm Re}(Y_1 + Y_2)\\ {\rm Im}(Y_1 + Y_2)\\ q_2 - q_1\end{pmatrix}\; .
\label{SE:BlochChoiRel}\end{align}
Again, it is easy to verify that the relation between this Bloch vector 
representation and the Choi representation is invertible.

\subsection{\label{SC} Criteria for positivity and complete positivity}

\paragraph{Local complete positivity} 
In order to verify if the Choi-matrix (as a linear transformation) 
projected onto the orthogonal subspace of $|\phi_{\rm B}\ra\la\phi_{\rm B}|$, 
is positive, we choose the orthonormal states
\begin{align}
|\psi_1\ra &= \frac{1}{\sqrt{2}}\; 
             \begin{bmatrix} 1\\ 0\\ 0\\ -1\end{bmatrix}\; , \quad
|\psi_2\ra = \begin{bmatrix} 0\\ 0\\ 1\\  0\end{bmatrix}\; , \quad
|\psi_3\ra = \begin{bmatrix} 0\\ 1\\ 0\\  0\end{bmatrix}\; ,
\label{SC:OrtSubBas}\end{align}
to span that subspace. Then we obtain for the matrix representation of the 
Choi matrix of $\mathcal{L}_t$, projected on that subspace:
\begin{align}
C_\perp &= \begin{pmatrix}
  {\rm Re}(Z_1) -\, \frac{q_1+q_2}{2} & \frac{X^* - Y_2  }{\sqrt{2}} & 
                                        \frac{X   + Y_1^*}{\sqrt{2}}        \\
  \frac{X   - Y_2^*}{\sqrt{2}}        & q_2                          & Z_2^*\\
  \frac{X^* + Y_1  }{\sqrt{2}}        & Z_2                          & q_1
\end{pmatrix}            \notag\\
%C_\perp &= \begin{pmatrix}
% {\rm Re}(Z_1) -\, \frac{q_1+q_2}{2} & \frac{X + Y_1^*}{\sqrt{2}} & 
%                                             \frac{X^* - Y_2}{\sqrt{2}} \\
% \frac{X^* + Y_1}{\sqrt{2}} & q_1 & Z_2 \\
% \frac{X - Y_2^*}{\sqrt{2}} & Z_2^* & q_2\end{pmatrix} \notag\\
&= \begin{pmatrix}
     D_{11} & D_{12} & D_{13} \\
     D_{21} & D_{22} & D_{23} \\
     D_{31} & D_{32} & D_{33} \end{pmatrix} \; .
%\begin{pmatrix}
%      D_{11} & D_{13} & D_{12} \\
%      D_{31} & D_{33} & D_{32} \\
%      D_{21} & D_{23} & D_{22} \end{pmatrix} \; .
\label{SC:LindMats}\end{align}
The second equality is obtained by solving Eq.~(\ref{SE:GenParams}) for the 
matrix elements $D_{ij}$. It simply means that $C_\perp = D$. 

We may now use the Sylvester criterion to check whether $D\ge 0$ or not.
A general discussion of that criterion can be found in the text 
book~\cite{Meyer00}; the present positive semidefinite case has been treated
in Ref.~\cite{Pru86}. In that case, the statement is the following:
A Hermitian matrix is positive semidefinite if and only if all principal minors
are larger or equal to zero. Hence, for $D\ge 0$, it must hold:
\begin{align}
&\hspace{14pt} D_{11}, D_{22}, D_{33} \ge 0\; , \quad 
D_{11} D_{33} - |D_{31}|^2 \ge 0\; , \notag\\
&D_{11} D_{22} - |D_{21}|^2 \ge 0\; , \quad D_{33} D_{22} - |D_{32}|^2 \ge 0
\; , \notag\\
&D_{11} D_{22} D_{33} + 2\, {\rm Re}(D_{12} D_{23} D_{31}) \ge\notag\\
&\qquad\qquad\qquad D_{11} |D_{32}|^2 + D_{22} |D_{31}|^2 + D_{33} |D_{21}|^2
\; .
\label{SC:ComplPosCond}\end{align}

\paragraph{Local positivity}
According to the criterion in Eq.~(\ref{CM:localP}), we need to verify that 
$\la\psi|\, \mathcal{L}[\, |\phi\ra\la\phi|\, ]\, |\psi\ra \ge 0$ for all 
$|\psi\ra \perp |\phi\ra$. Such general orthonormal states may be written 
as the column vectors of a unitary matrix, taken from the group $SU(2)$. 
Removing an ineffective global phase we find:
\[ |\psi\ra = \begin{pmatrix} \cos(\theta/2)\\ 
      \rme^{\rmi\beta}\, \sin(\theta/2)\end{pmatrix} \; , \qquad
   |\phi\ra = \begin{pmatrix} -\, \sin(\theta/2)\\
      \rme^{\rmi\beta}\, \cos(\theta/2)\end{pmatrix} \; . \]
Hence, we consider
$p(\theta,\beta) = \la\psi|\, \mathcal{L}[\, |\phi\ra\la\phi|\, ]\, |\psi\ra$
as a function of $\theta$ and $\beta$. Therefore, we may say that the 
$\mathcal{L}_t$ is positive at time $t$, if and only if $p(\theta,\beta) \ge 0$
for all $\theta$ and $\beta$. Using the parametrization of 
Eq.~(\ref{SE:ChoiGen}), $p(\theta,\beta)$ may be written as
\begin{align}
p(\theta,\beta) &=  \frac{q_1 + q_2}{2}\, \cos^2\theta 
     + \frac{q_2 - q_1}{2}\, \cos\theta + \frac{A}{2}\; \sin^2\theta \notag\\
&\quad
+ \frac{{\rm Re}\big [\, (Y_1 + Y_2)\, \rme^{-\rmi\beta}\, \big ]}{2}\;
   \sin\theta \\
&\quad + \frac{{\rm Re}\big [\, (Y_2 - Y_1)\, \rme^{-\rmi\beta} 
   - 2X\, \rme^{\rmi\beta}\, \big ]}{2}\; \sin\theta\; \cos\theta\; ,
\notag\end{align}
where $A= {\rm Re}[Z_1 - Z_2\, \rme^{-2\rmi\beta}]$.
In terms of the master equation parameters, we find
\begin{align}
&R= D_{22} + D_{33}\; , \quad
Y_1 + Y_2 = \sqrt{2}\, (D_{21} - D_{13})\; , \notag\\
&S= D_{33} - D_{22}\; , \quad 
A_1= D_{11} - \frac{D_{33} + D_{22}}{2} - {\rm Re}\, D_{23} \; , \notag\\
&Y_2 - Y_1 - 2\, X^* = -\, \sqrt{2}\, (D_{21} + D_{13})\; , 
\end{align}
such that
\begin{align}
2\, p(\theta,\beta)&= R + S\, \cos\theta
   + \Big ( D_{11} - \frac{R}{2}\Big )\, \sin^2\theta \notag\\
   {} &+ {\rm Re}\big [ - D_{23}\, \rme^{-2\rmi\beta}\, \sin\theta
    + \sqrt{2}\, (D_{21} - D_{13})\, \rme^{-\rmi\beta} \notag\\
   {} & - \sqrt{2}\, (D_{21} 
   + D_{13})\, \rme^{-\rmi\beta}\, \cos\theta\, \big ]\; \sin\theta \; .
\label{SP:ContrCond}\end{align}
This shows that positivity, just as complete positivity, only depends on 
the dissipation matrix $D$. 

In general, one should try to find all minima of this function and verify that 
those are non-negative. Since the domain of $p(\theta,\beta)$ is a torus 
without boundaries, it is sufficient to find the critical points where the 
partial derivatives $\partial p/\partial\theta$ and $\partial p/\partial\beta$ 
are both equal to zero. The corresponding equations may be reduced to a 
root-finding problem for 4'th order polynomials. Thus analytical expressions 
may be obtained in principle, even so they are probably not very useful. Still, 
numerical evaluations are pretty straight forward to implement. In Sec.~\ref{E} 
we will discuss different classes of generators, where particularly simple 
analytical solutions can be found.

\section{\label{E} Examples}

In this section, we consider three different classes of generators. For each 
class, the set of positive (completely positive) generators is interpreted as a 
region in a certain parameter space (a subspace of the $9$-dimensional vector
space of dissipation matrices). In general, these regions must be convex,
since the respective criteria involve expectation values of some linear matrix 
which represents the generator. Hence, if we consider the expectation value of 
any convex combination of two generators, it immediately decomposes into the 
corresponding convex combination of expectation values. Unless stated 
otherwise, we analyze the criteria for positivity and complete positivity in
terms of the dissipation matrix $D$.

\subsection{\label{EX} $X$-shaped quantum channels and generators}

The term ``$X$-shape'' refers to the case, where the non-zero elements in the 
Choi matrix appear to form the letter ``X'', that means that 
$Y_1 = Y_2 = X = 0$ in Eq.~(\ref{SE:ChoiGen}). Hence,
\begin{align}
C_\mathcal{L} = \begin{pmatrix}
   -q_1 & 0 & 0 & -Z_1^*\\
   0 & q_1 & Z_2 & 0\\
   0 & Z_2^* & q_2 & 0 \\
   -Z_1 & 0 & 0 & -q_2\end{pmatrix} \; .
\end{align}
In this case, the $X$-shape of the generator implies the $X$-shape of the 
quantum channel, and vice versa. Many important models lead to quantum channels
of that type~\cite{BrLaPi09,RHP14}. In terms of the Bloch vector 
representation, the $X$-shape implies that the dynamics along the $z$-axis is 
independent from that in the $(x,y)$-plane~\cite{Hall08}. 

According to Eq.~(\ref{SE:GenParams}) the $X$-shape of the Choi matrix 
$C_\mathcal{L}$ implies for $H$ and $D$ from the master equation representation 
in Eq.~(\ref{CP:LindMastEq}): $H_{12} = 0$, $D_{13}= D_{12}= 0$ as well as
\begin{align}
q_1 &= D_{22}\; , \;\qquad q_2 = D_{33}\; , \;\quad Z_2 = D_{23} \notag\\
\text{and}\quad 
Z_1 &= \rmi\, (H_{22} - H_{11}) + D_{11} + \frac{D_{33} + D_{22}}{2}\; .
\end{align}
For the matrix $C_\perp$ we thus obtain:
\begin{align}
C_\perp= \begin{pmatrix} D_{11} & 0 & 0\\
      0 & D_{22} & D_{23} \\
      0 & D_{32} & D_{33}\end{pmatrix}\; .
\label{EX:LindMats}\end{align}

\paragraph{Complete positivity}
Considering all principal minors of the dissipation matrix in 
Eq.~(\ref{EX:LindMats}), we find
\begin{align}
D_{11}, D_{22}, D_{33} &\ge 0\; ,&
D_{22}\, D_{33} - |D_{23}|^2 &\ge 0\; , \notag\\
D_{11}\, D_{22} &\ge 0\; ,& 
D_{11}\, \big [\, D_{22}\, D_{33} - |D_{23}|^2\, \big ] &\ge 0\; .
\end{align}
Removing redundant inequalities, we are left with
\begin{align}
D_{11}, D_{22}, D_{33} \ge 0\; , \quad D_{22}\, D_{33} \ge |D_{23}|^2\; .
\label{EX:CPosCond}\end{align}

\paragraph{Positivity}
From Eq.~(\ref{SP:ContrCond}) we find:
\begin{align}
2 p(\theta,\beta) &= R + S\; \cos\theta + A\; \sin^2\theta \ge 0 \; , \\
\text{where}\quad A&=  D_{11} - \frac{R}{2} 
   - {\rm Re}\big [\, D_{23}\, \rme^{-2\rmi\beta} \,\big ]\; ,
\notag\end{align}
$R= D_{22} + D_{33}$, and $S= D_{33} - D_{22}$.
This inequality must hold for all values of $\theta$ and $\beta$, parametrizing
the quantum state to which the generator is applied. Thus, we only need to
verify if the minimum of this expression is larger than zero. As far as $\beta$ 
is concerned, this means that we may replace $A$ by its minimum (as a function 
of $\beta$), which is given by $A_{\rm min} = D_{11} - R/2 - |D_{23}|$. We are
then left with the condition
\begin{align}
\forall\, \theta \;\; :\;\; R + S\; \cos\theta 
  + A_{\rm min}\; \sin^2\theta \ge 0\; .
\label{EX:thetamin}\end{align}
This condition is further evaluated in~\ref{aEX}. As a result, we find that 
the conditions for positivity become
\begin{align}
D_{22}, D_{33} \ge 0\; , 
\end{align}
and if $D_{11} < |D_{23}|$, in addition
\begin{align}
 \big |\, |D_{23}| - D_{11} \big | \le \sqrt{D_{22}\, D_{33}} \; .
\end{align}

\begin{figure}
\centering
\includegraphics[width=0.8\columnwidth]{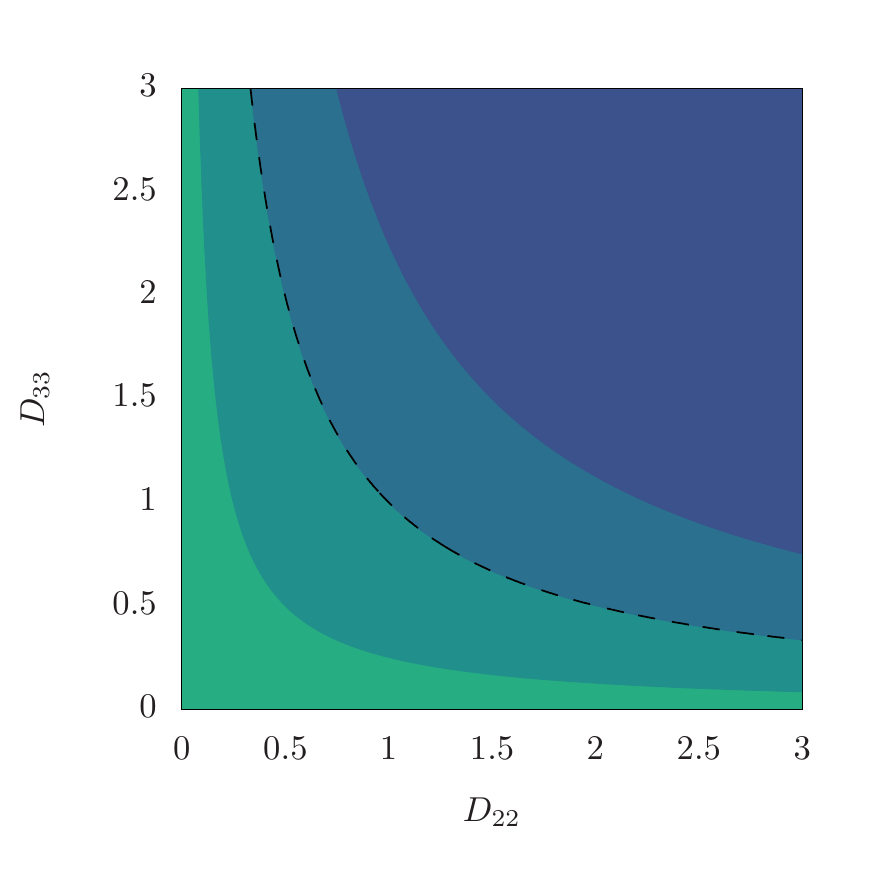}
\caption{
The parameter space $D_{22}, D_{33} \ge 0$ for visualizing the regions
of positivity and complete positivity for the $X$-shaped generator, for 
$|D_{32}| = 1$. For complete positivity, the point $(D_{22}, D_{33})$ must lie
above the black dashed line, while $D_{11} \ge 0$ is required. For positivity,
the allowed region for $(D_{22}, D_{33})$ depends on $D_{11}$: For 
$D_{11} \ge 1$, it is the whole quadrant; for $D_{11} = 1/2$ the allowed region 
consists of the dark green an blue areas; for $D_{11} = 0$ it consists of the 
blue areas above the black dashed line; and for $D_{11} = -1/2$ it consists of 
the dark blue area alone.}
\label{f:EX:scheme}\end{figure}

In Fig.~\ref{f:EX:scheme}, we show the parameter space $D_{22}, D_{33} \ge 0$ 
for visualizing the regions of positivity and complete positivity for the
$X$-shaped generator. For complete positivity, the inequalities to fulfill are 
given in Eq.~(\ref{EX:CPosCond}), which states independent conditions on 
$D_{11}$ on the one hand and $D_{22}, D_{33}, |D_{23}|^2$ on the other. For 
positivity, by contrast, the conditions on $D_{22}$ and  $D_{33}$ depend on 
$D_{11}$. Here, we observe an interesting behavior: As $D_{11}$ approaches zero 
from above, the region of positivity becomes more and more similar to the 
region of complete positivity, until they coincide for $D_{11} = 0$. When 
$D_{11}$ becomes negative, complete positivity is violated while positivity is 
still maintained sufficiently far away from the black dashed line.

\subsection{\label{EO} $O$-shaped quantum channels}

Here, we consider another subset of single qubit generators, which also allow 
for an analytic solution. These are in some sense complementary to the 
$X$-shaped channels, These are obtained from the general case by setting 
$q_1 = q_2 = q$, $Y_1 = - Y_2 = Y$, and $Z_2 = 0$. The Choi matrix, representing
the generator resembles an $O$, instead of an $X$, that is why we call them 
$O$-shaped channels. 
\begin{align}
C_{\mathcal{L}} = \begin{bmatrix} -q & Y^* & X^* & -Z_1^*\\
      Y & q & 0 & -X^*\\
      X & 0 & q & -Y^*\\
      -Z_1 & -X & -Y & -q\end{bmatrix}
\end{align}
According to Eq.~(\ref{SE:GenParams}), this implies
for the matrix elements of $H$ and $D$ from the master 
equation~(\ref{CP:LindMastEq}):
%\begin{align}
%&q= D_{33}\; , \quad Z_1= \rmi\, (H_{22}-H_{11}) + D_{11} + D_{33}\; , \notag\\
%&Y= -\rmi\, H_{21} + \frac{D_{13}}{\sqrt{2}}\; , \quad
%X^*= \rmi\, H_{21} + \frac{D_{13}}{\sqrt{2}}\; ,
%\end{align}
\begin{align}
q&= D_{33}\; , \qquad Z_1= \rmi\, (H_{22}-H_{11}) + D_{11} + D_{33}\; , \notag\\
Y&= -\rmi\, H_{21} + \frac{D_{13}}{\sqrt{2}}\; , \quad
X^*= \rmi\, H_{21} + \frac{D_{13}}{\sqrt{2}}\; ,
\end{align}
with $D_{22}= D_{33}$, $D_{21}= D_{13}$, and $D_{23} = 0$. The matrix for 
verifying complete positivity reads
\begin{align}
C_\perp = \begin{pmatrix} D_{11} & D_{31} & D_{13}\\
      D_{13} & D_{22} & 0\\ D_{31} & 0 & D_{22}\end{pmatrix} \; ,
\end{align}

\paragraph{Complete positivity}
Expressed in terms of the dissipation matrix, considering all principal minors.
%\begin{align}
%&D_{11}, D_{22} \ge 0\; , \quad D_{11} D_{22} - |D_{13}|^2 \ge 0
%\; , \notag\\
%&D_{11}\, D_{22}^2 - D_{13} D_{31} D_{22} - D_{31} D_{22} D_{13} \ge 0
%\end{align}
\begin{align}
      D_{11}, D_{22} \ge 0\; , \quad D_{11} D_{22} - |D_{13}|^2 &\ge 0
\; , \notag\\
D_{11}\, D_{22}^2 - D_{13} D_{31} D_{22} - D_{31} D_{22} D_{13} &\ge 0
\end{align}
This can be reduced to
\begin{align}
D_{11}, D_{22} \ge 0\; , \quad D_{11} D_{22} \ge 2\, |D_{13}|^2 \; . 
\label{EO:CPosCond}\end{align}

\paragraph{Positivity}
Under the conditions mentioned above, the function $2\, p(\theta,\beta)$ from
Eq.~(\ref{SP:ContrCond}) becomes (in terms of the master equation parameters)
\begin{align}
2p(\theta,\beta)&=  2\, D_{22} + (D_{11} - D_{22})\, \sin^2\theta \notag\\
&-2\sqrt{2}\, {\rm Re}\big [ D_{13}\, \rme^{-\rmi\beta}\big ]\;
             \sin\theta\cos\theta \geq 0 \; .
\label{EO:PosCond}\end{align}
Again, it is possible to derive the conditions for positivity, which do no 
longer involve the angles $\theta$ and $\beta$. The respective calculation is 
outlined in~\ref{aEO}, with the result [see Eq.~(\ref{aEO:result})]
\begin{align}
3\, D_{22} + D_{11} \ge 0 \; , \quad
D_{22}\; (D_{11} + D_{22}) \ge |D_{13}|^2\; .
\label{EO:PosCondRes}\end{align}

\begin{figure}
\centering
\includegraphics[width=0.8\columnwidth]{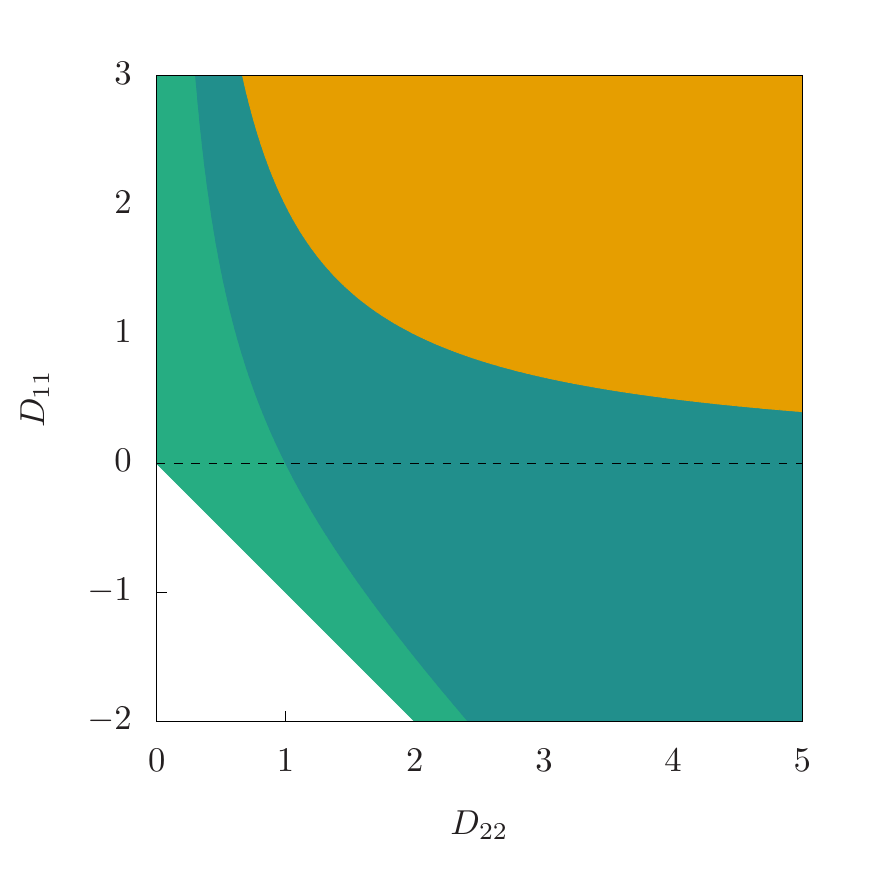}
\caption{The parameter space $D_{22}, D_{11}$ for visualizing the regions of 
positivity [Eq.~(\ref{EO:PosCondRes})] and complete positivity 
[Eq.~(\ref{EO:CPosCond})] for the $O$-shaped generator. For $|D_{13}| = 0$, the 
region of complete positivity is simply the positive quadrant 
$D_{22}, D_{11} \ge 0$, while the region for positivity is the whole colored 
region. For $|D_{13}| = 1$, the region of complete positivity is colored in 
orange, the region of positivity is dark green and orange.}
\label{f:EO:scheme}\end{figure}

In Fig.~\ref{f:EO:scheme}, we show the different regions of positivity and
complete positivity in the parameter space of $D_{22}, D_{11}$. We distinguish
two qualitatively different cases, $|D_{13}| = 0$ and $|D_{13}| = 1$. In both
cases, the region of positivity is considerably larger than the region for 
complete positivity.

\subsection{\label{EP} Non-unital anisotropic Pauli channels}

Here, $\Lambda_t$ is given as an affine transformation of state vectors
in the Bloch sphere~\cite{BenZyc06} (see the corresponding paragraph in 
Sec.~\ref{SE}): 
\begin{align}
\Lambda_t\;\; :\;\; \vec v \to \vec v' = R\; \vec v + \vec s \; ,
\end{align}
where $R$ is a real diagonal matrix and $\vec s$ a real vector.
\begin{align}
R = \begin{pmatrix} R_{11} & 0 & 0\\ 0 & R_{22} & 0\\ 0 & 0 & R_{33}
    \end{pmatrix} \; ,
\qquad \vec s = \begin{pmatrix} s_1\\ s_2\\ s_3\end{pmatrix}\; .
\end{align}
Using the general formula, Eq.~(\ref{CP:Generator}), for constructing the 
generator, we find
\begin{align}
\mathcal{L}_{\rm P} \;\; :\;\; \vec v \to \vec v' = \frac{\rmd R}{\rmd t}\;
   \big [\, R^{-1}\, (\vec v - \vec s)\, \big ] + \frac{\rmd \vec s}{\rmd t}
   \; . 
\label{EP:PauliLt}\end{align}
Hence, the generator for this Pauli channel is given by the affine 
transformation $\vec v \to \vec v' 
   = R_{\rm \mathcal{L}P}\; \vec v + \vec t_{\rm \mathcal{L}P}$, with
\begin{align}
   R_{\rm \mathcal{L}P} = \begin{pmatrix} -\gamma_1 & 0 & 0\\
      0 & -\gamma_2 & 0\\ 0 & 0 & -\gamma_3\end{pmatrix} \; , \quad
   t_{\rm \mathcal{L}P} = \begin{pmatrix} \tau_1\\ \tau_2\\ \tau_3\end{pmatrix} 
\; , \notag\\
\text{where}\quad 
 \gamma_j = \frac{-1}{R_{jj}}\, \frac{\rmd R_{jj}}{\rmd t}\; , \qquad
   \tau_j = \frac{\rmd\, t_j}{\rmd t} + \gamma_j\; t_j \; . 
\end{align}
The Choi matrix representation of $\mathcal{L}_{\rm P}$ is obtained by 
inverting Eq.~(\ref{SE:BlochChoiRel}), with the result
\begin{align}
C_{\rm \mathcal{L}{\rm P}} = \frac{1}{2}\begin{pmatrix}
   -\gamma_3 \! +\!  \tau_3\! & \tau_1 \! -\! \rmi \tau_2 & 0 & -\gamma_1 \! -\!  \gamma_2\\
   \tau_1 \! +\! \rmi \tau_2 & \gamma_3 \! -\!  \tau_3 & \gamma_2 \! -\!  \gamma_1\! & 0\\
   0 & \gamma_2 \! -\!  \gamma_1 & \gamma_3 \! +\!  \tau_3\! & \tau_1 \! -\! \rmi \tau_2\\
   -\gamma_1 \! -\!  \gamma_2\! & 0 & \tau_1 \! +\!  \rmi \tau_2\! & -\gamma_3 \! -\!  \tau_3
 \end{pmatrix}\; .
\end{align}
In what follows, we compute the positivity and the complete positivity
condition in terms of the parameters $\gamma_j$ and $\tau_j$, since this allows
for relatively simple geometric interpretations. For the parametrization in 
terms of the master equation~(\ref{CP:LindMastEq}), we obtain from 
Eq.~(\ref{SE:ChoiGen}) and~(\ref{SE:GenParams}):
\begin{align}
	D_{22}&= \frac{\gamma_3 - \tau_3}{2}\; ,&
	D_{33}&= \frac{\gamma_3 + \tau_3}{2}\; ,&
	H_{22}&= H_{11}\; , \notag\\
	D_{11}&= \frac{\gamma_1 + \gamma_2 - \gamma_3}{2}\; ,&
	D_{23}&= \frac{\gamma_2 - \gamma_1}{2}\;,& H_{12}&= 0\; , \notag\\
	D_{21}&= \frac{\tau_1 + \rmi\tau_2}{2\sqrt{2}} = -D_{13}\; .
\label{EP:PauPars}\end{align}
This yields
\begin{align}
&C_\perp = \frac{1}{2}\begin{pmatrix} 
   \gamma_1 + \gamma_2 - \gamma_3 & w^*  & -w\\
       w & \gamma_3 - \tau_3 & \gamma_2 - \gamma_1 \\
        -w^* & \gamma_2 - \gamma_1 & \gamma_3 + \tau_3\end{pmatrix}\; ,
\label{EP:PauCperp}\end{align}
with $w= (\tau_1 +\rmi \tau_2)/\sqrt{2}$.

\begin{figure}
\centering
\includegraphics[width=0.8\columnwidth]{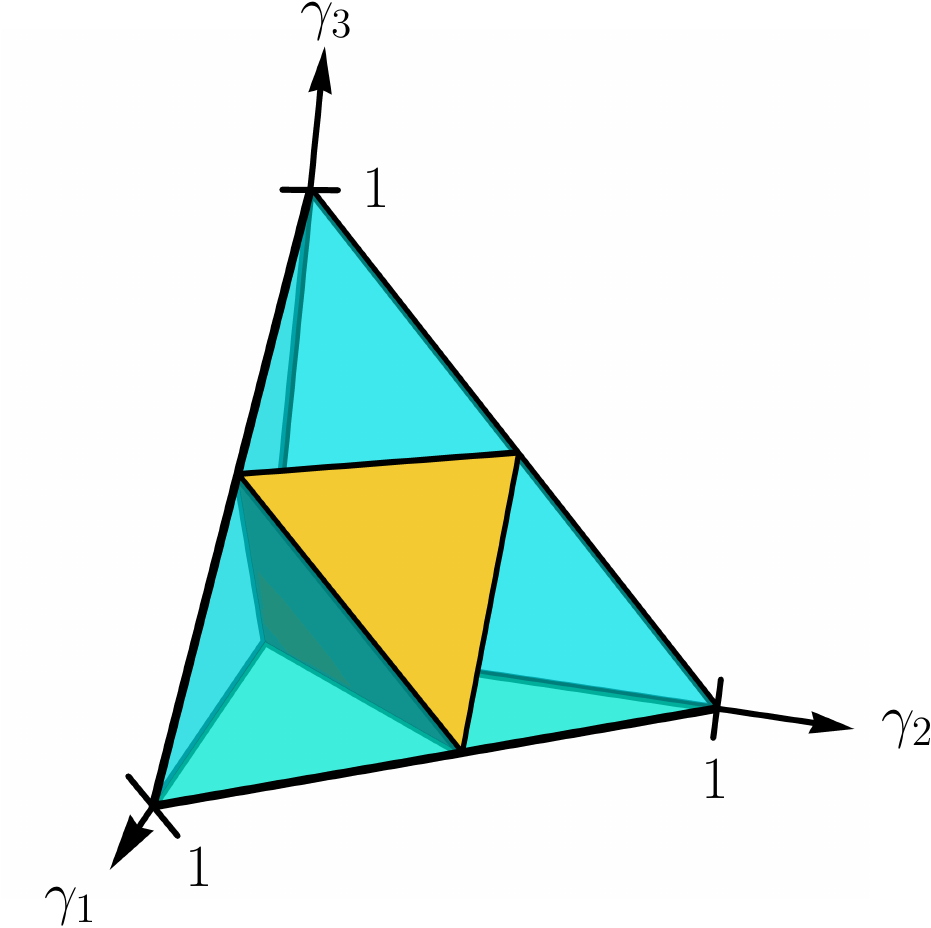}
\caption{Parameter space of $\gamma_1$, $\gamma_2$ and $\gamma_3$. For the 
generator $\mathcal{L}_{\rm P}$ in Eq.~(\ref{EP:PauliLt}) to be positive, all 
elements $\gamma_j$ must be positive (blue transparent color). For 
$\mathcal{L}_{\rm P}$ to be completely positive, the elements $\gamma_j$ must 
fulfill the conditions in Eq.~(\ref{EP:gammaCP}). The corresponding region is 
colored in orange. Note that we show a cut through the regions of positivity 
and complete positivity which really extend towards arbitrary large positive 
values. }
\label{f:EP:scheme}\end{figure}

\begin{figure}
\centering
\includegraphics[width=0.8\columnwidth]{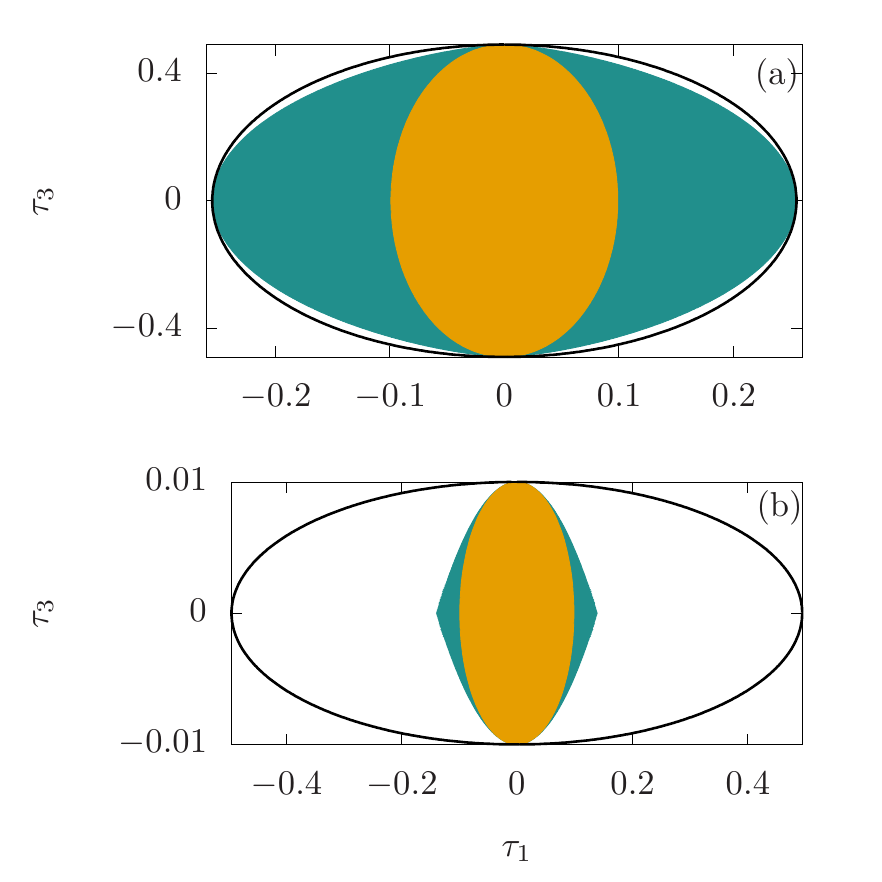}
\caption{Comparison of the region of positivity and complete positivity in the
 parameter space for $\vec\tau$ for $\gamma_1= \gamma_2$. Black solid line
shows the ellipsoid $\vec\tau = \underline\gamma\, \vec\rme_r$, the orange 
region shows the region of complete positivity, the green region (including
orange) the region of positivity. In panel (a), 
$\gamma_1 = 0.255, \gamma_3 = 0.49$ which amounts to an ellipsoid of the shape
of an rugby ball, in panel (b) $\gamma_1= 0.495, \gamma_3= 0.01$ where
the ellipsoid looks more like a flat pancake.}
\label{f:Pauli}\end{figure}

\paragraph{Complete positivity}
The complete derivation can be found in~\ref{aEP}. It yields separate 
conditions for the diagonal elements $\gamma_j$ and the vector $\vec\tau$.
For the diagonal elements $\gamma_j$ we find:
\begin{align}
&\forall\; i\ne j\ne k\ne i \;\; :\;\;
  |\gamma_i - \gamma_j| \le \gamma_k \le \gamma_i + \gamma_j \; .
\label{EP:gammaCP}\end{align}
The corresponding region in the parameter space of the elements $\gamma_j$ is 
depicted as a orange region in Fig.~\ref{f:EP:scheme}. Assuming these 
conditions are fulfilled, the vector $\vec\tau$ must lie inside the following 
ellipsoid:
\begin{align}
&\frac{\tau_1^2}{a_1^2} + \frac{\tau_2^2}{a_2^2} + \frac{\tau_3^2}{a_3^2} \le 1
\; , \qquad a_1 = \gamma_1^2 - (\gamma_2 - \gamma_3)^2\; , \notag\\
& a_2 = \gamma_2^2 - (\gamma_1 - \gamma_3)^2\; , \quad
 a_3 = \gamma_3^2 - (\gamma_1 - \gamma_2)^2\; .
\end{align}
The regions of $\vec\tau$ where the generator $\mathcal{L}_{\rm P}$ fulfills 
the conditions of complete positivity are shown in Fig.~\ref{f:Pauli} in 
orange. Note that in this figure, we consider two particular cases, where 
$\gamma_1 = \gamma_2$ such that the resulting ellipsoid as defined above is
symmetric with respect to the $\tau_3$ axis.

\paragraph{Positivity}
In the general expression for $p(\theta,\beta)$ in Eq.~(\ref{SP:ContrCond}), we 
replace the parameters with those from the Pauli channel, given in
Eq.~(\ref{EP:PauPars}). This yields
\begin{align}
2\, p(\theta,\beta)  
&= \gamma_3\; \cos^2\theta + \big [\, 
   \gamma_1\; \cos^2\beta + \gamma_2\; \sin^2\beta \, \big ]\; \sin^2\theta
\notag\\
&+ \tau_3\; \cos\theta + \big (\, \tau_1\; \cos\beta 
   + \tau_2\; \sin\beta\, \big )\; \sin\theta\; .
\end{align}
We can express the general inequality $2\, p(\theta,\beta) \ge 0$ in a
geometric form:
\begin{align}
\vec\rme_r = \begin{pmatrix} \sin\theta\, \cos\beta\\
   \sin\theta\, \sin\beta\\
   \cos\theta\end{pmatrix} \quad :\quad \vec\rme_r \cdot \big ( 
      \underline{\gamma}\, \vec\rme_r + \vec\tau  \big ) \ge 0 \; ,
\label{EP:geomtauinequal}\end{align}
where $\underline{\gamma}$ is the diagonal matrix with elements $\gamma_j$.

The interpretation of this result is easy: 
$-\underline\gamma\, \vec\rme_r + \vec\tau$ is the image of $\vec\rme_r$ under
the generator $\mathcal{L}_{\rm P}$. Thus an infinitesimal intermediate map
would yield
\[ \Lambda_{t,t+\delta} \;\; :\;\; \vec\rme_r \to \vec\rme_r{}' = \vec\rme_r
+ \delta\; \mathcal{L}_{\rm P}[\vec\rme_r] \; . \]
In order to have $\|\vec\rme_r{}' \| \le 1$, the image under the generator must 
be pointing towards the center of the Bloch sphere, i.e. the scalar product
between $-\underline\gamma\, \vec\rme_r + \vec\tau$ and $\vec\rme_r$ must be 
negative. Multiplying the resulting inequality by minus one, we find
\[ \forall\vec\rme_r \;\; :\;\; \vec\rme_r \cdot \big ( 
      \underline{\gamma}\, \vec\rme_r - \vec\tau  \big ) \ge 0 \; . \]
This relation is equivalent to the inequality in Eq.~(\ref{EP:geomtauinequal}),
as can be seen by replacing $\vec\rme_r$ by $-\vec\rme_r$. 

As shown in~\ref{aPP}, the set of $\vec\tau$ for which the 
Pauli generator $\mathcal{L}_{\rm P}$ is positive, i.e. the inequality in 
Eq.~(\ref{EP:geomtauinequal}) holds, is the convex region, which contains the
origin and is limited by the surface [see Eq.~(\ref{aPP:finres})] 
\begin{align}
\mathcal{T} = \{ \vec\tau(\theta,\beta) = (\vec\rme_r \cdot \underline\gamma\,
   \vec\rme_r) \; \vec\rme_r - 2\, \underline{\gamma}\vec\rme_r \, \} \; .
\end{align}
In Fig.~\ref{f:Pauli}, we show the region in $\vec\tau$-space which corresponds
to positivity and complete positivity of the Pauli channel generator 
$\mathcal{L}_{\rm P}$. We consider two cases:
$\gamma_1 = \gamma_2 = 0.255$, $\gamma_3 = 0.49$ in panel (a), and
$\gamma_1 = \gamma_2 = 0.495$, $\gamma_3 = 0.01$ in panel (b). In the yellow
triangle shown in Fig.~\ref{f:EP:scheme}, these points are located near the 
upper horizontal line (a) and near the lower corner (b), respectively. Choosing
$\gamma_1 = \gamma_2$ leads to regions of (complete) positivity, which are 
symmetric with respect to the $\tau_3$-axis, which allows us to show 
two-dimensional projections. We find that the regions of positivity and complete
positivity are always contained in ellipsoid with the parametrization
$\vec\tau(\theta,\beta) = \underline{\gamma}\, \vec\rme_r$. As required, the
region of complete positivity (orange) is fully contained in the region of 
positivity (olive green). In panel (a), we show a case where the ellipsoid
$\underline{\gamma}\, \vec\rme_r$ resemble roughly a rugby ball. In that case,
the are only rather thin stripes near the border of the ellipsoid, where the 
generator is not positive any more. In panel (b), the ellipsoid has the shape
of a flat pancake, and the region of positivity in the center is much smaller.

\section{\label{F} Conclusions}

In order to determine whether a given differentiable quantum process is 
CP-divisible and/or P-divisible, we derive criteria which can be applied to the
generator of the process. For the single qubit case, we discuss three common
representations of the generator and work out the one-to-one mappings between 
them. We find criteria for CP- and P-divisibility, which can be expressed as 
inequalities in terms of the elements of the dissipation matrix. In the
CP case, we avoid solving an eigenvalue problem by using the principal minor
test for semidefinite matrices. In the P case, the corresponding inequality
must be fulfilled for a whole two-parameter family of functions, which leads to
an optimization problem without explicit general solution.

We then discuss three different classes of generators, where our criteria do
yield explicit results for CP- and P-divisibility: the familiar $X$-shaped 
channels where the elements of the Choi matrix are non-zero in the diagonal and
the anti-diagonal, only; the so called $O$-shaped channels, where $C_{23} = 0$, 
$C_{11} = C_{44}$ and $C_{12} = -C_{34}$; and most importantly the non-unital 
Pauli channels.

Besides its general value, as for instance the positivity criteria for the 
Pauli channel, we expect our results to prove useful in the area of quantum
process tomography and the construction of optimal P-divisible or CP-divisible
approximations to non-Markovian quantum processes. In particular there, the 
renouncement on the calculation of higher order roots may help to find 
analytical or semi-analytical solutions.

\section*{Acknowledgements}
We gratefully acknowledge Sergey Filippov for useful discussions, as well as
Carlos Pineda for valuable comments on the manuscript.

\begin{appendix}
\section{\label{aEX} Positivity of $X$-shaped generators}

The condition in Eq.~(\ref{EX:thetamin}) can be expressed equivalently in terms 
of the variable $x = \cos\theta$ as follows:
\begin{align}
&\forall\, x\in [-1,1] \;\; :\notag\\
&f(x)= (R - A_{\rm min})\; x^2 + S\; x + A_{\rm min} \ge 0 \; .
\end{align}
First note the following obviously necessary conditions
\begin{align}
f(0) \;\; &:\;\; A_{\rm min} \ge 0\quad\text{and}\notag\\
f(\pm 1) \;\; &:\;\; R \pm S \ge 0 \quad\Leftrightarrow\quad 
0 \le |S| \le R \; .
\label{XP:poscond}\end{align}
To find the necessary and sufficient conditions, we will divide the problem in
two cases: (i) $\Delta = R - A_{\rm min} \le 0$ and (ii) $\Delta > 0$.

In case (i) the conditions in Eq.~(\ref{XP:poscond}) are also sufficient as can
be seen as follows: $f(x)$ is convex, such that for any $x_1,x_2$ and
$0 < \lambda < 1$:
\[ f(\lambda\, x_1 + (1-\lambda)\, x_2) \ge \lambda\; f(x_1) + (1-\lambda)\; 
      f(x_2) \; . \]
Choosing $x_1 = -1$ and $x_2 = 1$, we find
\[ f(1-2\lambda) \ge \lambda\; f(-1) + (1-\lambda)\; f(1) \; , \]
which implies that $f(x) \ge 0 $ in the interval $(-1,1)$.

In case (ii) $\Delta > 0$, the conditions in Eq.~(\ref{XP:poscond}) are not
sufficient. In this case, positivity requires that either $f(x)$ has no zeros,
or its zeros
\[ x_{1,2} = -\, \frac{S}{2\Delta} \pm \sqrt{\frac{S^2}{\Delta^2} - \frac{4\, A_{\rm min}}{\Delta}} \; , \]
are lying both to the left or both to the right of the interval $(-1.1)$. This
can be expressed as
\[ |S| \le 2\, \sqrt{A_{\rm min}\, \Delta} \quad\text{or}\quad
   |S| \ge 2\Delta + \sqrt{S^2 - 4 A_{\rm min}\, \Delta} \; . \]
The inequality to the right is equivalent to
\[ |S| \ge 2\Delta \quad\text{and}\quad 
   (\, |S| - 2\Delta\, )^2 \ge S^2 - 4 A_{\rm min}\, \Delta \; , \]
which is equivalent to
\[ |S| \ge 2\Delta \quad\text{and}\quad |S| \le R \; , \]
where $|S| \le R$ had already been identified as a necessary condition,
previously. Therefore, in case (ii) the necessary and sufficient conditions for
positivity are $A_{\rm min} \ge 0\; , \quad 0 \le |S| \le R$ and
\[   |S| \le 2\, \sqrt{A_{\rm min}\, \Delta} \quad\text{or}\quad
   |S| \ge 2\Delta \; . \]
It turns out that for $\Delta \le A_{\rm min}$ it holds that
$2\, \sqrt{A_{\rm min}\, \Delta} < 2\Delta$ such that the two conditions cancel 
each other, i.e. one of the two conditions is always fulfilled. For 
$\Delta > A_{\rm min}$ which is equivalent to $2\Delta > R$, by contrast, 
implies that $|S| \ge 2\Delta$ cannot hold, such that
$|S| \le 2\, \sqrt{A_{\rm min}\, \Delta}$ must be fulfilled.
To summarize, the necessary and sufficient conditions for positivity are as
follows:
\begin{itemize}
\item $A_{\rm min}= {\rm Re}\, Z_1 - |Z_2| \ge 0\; , \quad 0 \le |S| \le R$.
\item In addition, if $R > 2A_{\rm min} \quad :$\\
\[ |S| \le 2\, \sqrt{A_{\rm min}\, (R - A_{\rm min})} \; . \]
\end{itemize}
For the parametrization in terms of the master equation, we find that 
positivity only depends on the dissipation matrix $D$. Since
\begin{align}
R= D_{33} + D_{22}\; , \quad S= D_{22} - D_{33} \; ,
\end{align}
the condition $0\le |S| \le R$ implies that both $D_{22}$ and $D_{33}$ must be
larger than or equal to zero. Furthermore, with
\begin{align}
A_{\rm min} = D_{11} + \frac{D_{33} + D_{22}}{2} - |D_{32}|\; ,
\end{align}
the condition
$R > 2\, A_{\rm min}$ implies that $|D_{32}| > D_{11}$. Finally,
\begin{align}
	A_{\rm min} (R - A_{\rm min}) &= 
   \Big [ \frac{D_{33} + D_{22}}{2} - \big ( |D_{32}| - D_{11}\big ) \Big ] 
\times\notag\\
&\phantom{{}={}}
   \Big [ \frac{D_{33} + D_{22}}{2} + \big ( |D_{32}| - D_{11}\big ) \Big ]
\notag\\
&= \frac{(D_{33} + D_{22})^2}{4} - \big ( |D_{32}| - D_{11}\big )^2\; ,
\end{align}
such that $|S| \le 2\, \sqrt{A_{\rm min}\, (R - A_{\rm min})}$ is equivalent to
\begin{align}
&|D_{22} - D_{33}|^2 \le (D_{33} + D_{22})^2 
   - 4\, \big ( |D_{32}| - D_{11}\big )^2 \notag\\
&\Leftrightarrow\quad 
   \big ( |D_{32}| - D_{11}\big )^2 \le D_{33}\, D_{22} \; .
\end{align}
To summarize, in this parametrization, the conditions for positivity read
\begin{align}
D_{22}, D_{33} \ge 0\; , \quad 
D_{11} - |D_{32}| + \frac{D_{33} + D_{22}}{2} \ge 0
\label{aEX:A7}\end{align}
and if $D_{11} < |D_{32}|$, in addition
\begin{align}
 \big |\, |D_{32}| - D_{11} \big | \le \sqrt{D_{33}\, D_{22}} \; .
\end{align}
Note that this last inequality implies the second inequality of 
Eq.~(\ref{aEX:A7}), which can therefore be ignored.

\section{\label{aEO} Positivity of $O$-shaped generators}

We start from the condition for positivity in Eq.~(\ref{EO:PosCond}).
Using  the trigonometric identities $2\sin^2\theta=1-\cos2\theta$ and
$\sin2\theta=2\sin\theta\cos\theta$, Eq.~\eqref{EO:PosCond} becomes
\begin{align}
2 p(\theta, \beta) &= \frac{3 D_{22} + D_{11}}{2} 
   + \frac{D_{22} - D_{11}}{2}\; \cos\, 2\theta\notag\\
&\qquad\qquad 
   - \sqrt{2}\, {\rm Re}\big (\, D_{12}\, \rme^{-\rmi\beta}\, \big )\; 
   \sin\, 2\theta \ge 0 \; .
\end{align}
This expression is minimized with respect to $\beta$, simply by making sure
that ${\rm Re}(\, D_{12}\, \rme^{-\rmi\beta}\, ) = \pm\, |D_{12}|$. In other 
words: $2 p(\theta, \beta) \ge 0$ for all $\beta$ and $\theta$ is equivalent to
\begin{align}
\frac{3 D_{22} + D_{11}}{2} 
   + \frac{D_{22} - D_{11}}{2}\; \cos\, 2\theta
   \pm \sqrt{2}\, |D_{12}|\; \sin\, 2\theta \ge 0 \; .
\end{align}
This condition is equivalent to
\begin{align}
    \frac{3 D_{22} + D_{11}}{2} &\ge 0 \quad\text{and}\notag\\
\frac{(3 D_{22} + D_{11})^2}{4} &\ge \frac{(D_{22} - D_{11})^2}{4} 
   + 2 |D_{12}|^2
\end{align}
These two inequalities are equivalent to
\begin{align}
3 D_{22} + D_{11} \ge 0 \quad\text{and}\quad
D_{22}\, (D_{22} + D_{11}) \ge |D_{12}|^2
\label{aEO:result}\end{align}

\section{\label{aEP} Complete positivity of the non-unital anisotropic Pauli 
channel}

For the generator $\mathcal{L}_{\rm P}$ to be completely positive, the matrix 
$C_\perp$ given in Eq.~(\ref{EP:PauCperp}) must fulfill the inequalities in 
Eq.~(\ref{SC:ComplPosCond}). In the present case, this yields three sets of 
inequalities:
\begin{align}
\gamma_1 + \gamma_2 - \gamma_3 \ge 0\; , \quad \gamma_3 - \tau_3 &\ge 0\; , 
\quad \gamma_3 + \tau_3 \ge 0\; , \notag\\
(\gamma_3 - \tau_3) (\gamma_3 + \tau_3) - (\gamma_2 - \gamma_1)^2 &\ge 0 \; ,
\label{aEP:set1}\end{align}
\begin{align}
&(\gamma_1 + \gamma_2 - \gamma_3)(\gamma_3 - \tau_3) - |w|^2 \ge 0\; , \notag\\
&(\gamma_1 + \gamma_2 - \gamma_3)(\gamma_3 + \tau_3) - |w|^2 \ge 0\; , 
\label{aEP:tau12cond}\end{align}
and
\begin{align}
&(\gamma_1 + \gamma_2 - \gamma_3)\; \big [ 
   (\gamma_3 - \tau_3) (\gamma_3 + \tau_3) - (\gamma_2 - \gamma_1)^2 \big ]
\notag\\
&\qquad - w\;\;\, 
   \big [w^*\, (\gamma_3 + \tau_3) + w\, (\gamma_2 - \gamma_1)\big ] \notag\\
&\qquad - w^*\; 
   \big [ w^*\, (\gamma_2 - \gamma_1) + w\, (\gamma_3 - \tau_3) \big ] \ge 0
   \; ,
\end{align}
where $w= (\tau_1 + \rmi\, \tau_2)/\sqrt{2}$. From Eq.~(\ref{aEP:set1}), we 
find
\[ \gamma_3 \ge |\tau_3| \ge 0\; , \quad \gamma_1 + \gamma_2 \ge \gamma_3 
\; , \quad
(\gamma_2 - \gamma_1)^2 \le \gamma_3^2 - \tau_3^2 \; , \]
which yields the following conditions as necessary conditions (since we set
$\tau_3 = 0$ to arrive there):
\begin{align}
\gamma_1, \gamma_2, \gamma_3 \ge 0\; , \quad 
|\gamma_2 - \gamma_1| \le \gamma_3 \le \gamma_1 + \gamma_2 \; .
\end{align}
It is easy to verify that these inequalities are invariant under any 
permutation of indices; see Fig.~\ref{f:EP:scheme}. The remaining conditions,
may be interpreted as conditions for the vector $\vec\tau$. These consist of
the inequalities in Eq.~(\ref{aEP:tau12cond}) together with 
\begin{align}
&|\tau_3| \le \sqrt{\gamma_3^2 - (\gamma_2 - \gamma_1)^2}\; , 
\quad \text{and}\quad \label{aEP:tau3cond}\\
&(\gamma_1 + \gamma_2 - \gamma_3)\; (\gamma_2 + \gamma_3 - \gamma_1)\;
   (\gamma_3 + \gamma_1 - \gamma_2) \ge \notag\\
& (\gamma_1 + \gamma_2 - \gamma_3)\; \tau_3^2 
+ (\gamma_2 + \gamma_3 - \gamma_1)\; \tau_1^2
+ (\gamma_3 + \gamma_1 - \gamma_2)\; \tau_2^2 \; .
\label{aEP:tauFullCond}\end{align}
In~\ref{aEPO} we demonstrate that condition~(\ref{aEP:tauFullCond}) implies
all other conditions for the vector $\tau$, which can therefore be omitted.
Reorganizing the terms in Eq.~(\ref{aEP:tauFullCond}), we arrive at
\begin{align}
&\frac{\tau_1^2}{a_1^2} + \frac{\tau_2^2}{a_2^2} + \frac{\tau_3^2}{a_3^2} \le 1
\; , \qquad a_1 = \gamma_1^2 - (\gamma_2 - \gamma_3)^2\; , \notag\\
& a_2 = \gamma_2^2 - (\gamma_1 - \gamma_3)^2\; , \quad
 a_3 = \gamma_3^2 - (\gamma_1 - \gamma_2)^2\; .
\label{aEP:taucond}\end{align}

\subsection{\label{aEPO} Omissible inequalities for $\vec\tau$}

In what follows, we demonstrate that Eq.~(\ref{aEP:tau3cond}) as well as
Eq.~(\ref{aEP:tau12cond}) follow from Eq.~(\ref{aEP:taucond}) such that we may 
consider Eq.~(\ref{aEP:taucond}) as the only condition on $\vec\tau$. To that
end note first that setting $\tau_1 = \tau_2 = 0$ we can make the LHS of 
Eq.~(\ref{aEP:taucond}) only smaller which hence implies
\[ \tau_3^2 \le a_3^2 = \gamma_3^2 - (\gamma_1 - \gamma_2)^2 \; , \]
which is exactly Eq.~(\ref{aEP:tau3cond}). To show that Eq.~(\ref{aEP:taucond}) 
also implies Eq.~(\ref{aEP:tau12cond}), it is convenient to express $\vec\tau$ 
in elliptical coordinates,
\[ \vec\tau = \lambda\begin{pmatrix} a_1\; \sin\theta\, \cos\varphi\\
      a_2\; \sin\theta\, \sin\varphi\\
      a_3\; \cos\theta\end{pmatrix} \; , \]
such that Eq.~(\ref{aEP:taucond}) allows arbitrary values for the angles 
$\theta,\varphi$ and limits $\lambda$ to the range $0\le \lambda\le 1$. 

The two inequalities in Eq.~(\ref{aEP:tau12cond}) may be combined, and then
read
\[ \gamma_3 \pm \lambda\, a_3\, \cos\theta \ge \frac{\lambda^2\, \sin^2\theta\;
      (a_1^2\, \cos^2\varphi + a_2^2\, \sin^2\varphi)}
   {2\, (\gamma_1 + \gamma_2 - \gamma_3)}\; .  \]
Since $a_1^2$ and $a_2^2$ have the common factor 
$(\gamma_1 + \gamma_2 - \gamma_3)$ this inequality simplifies to 
\begin{align}
(\gamma_3 \pm \lambda\, a_3\, \cos\theta) &\ge 
      \frac{\lambda^2\, \sin^2\theta}{2}\; \big [
   (\gamma_3 + \gamma_1 - \gamma_2)\, \cos^2\varphi \notag\\
&\phantom{\ge 
      \frac{\lambda^2\, \sin^2\theta}{2}}
   + (\gamma_3 - \gamma_1 + \gamma_2)\, \sin^2\varphi \big ] \notag\\
&= \frac{\lambda^2\, \sin^2\theta}{2}\; \big [ \gamma_3 
      + (\gamma_1 - \gamma_2)\, \cos(2\varphi) \big ]
\end{align}
Due to the conditions in Eq.~(\ref{aEP:taucond}), we may assume that 
$\gamma_3 \ge a_3$ and $\gamma_3  \ge |\gamma_1 - \gamma_2|$. Therefore, in 
order to show that Eq.~(\ref{aEP:tau12cond}) holds, it is sufficient to prove
that
\[ \gamma_3 \pm \lambda\, a_3\, \cos\theta \ge 
   \frac{\lambda^2\, \sin^2\theta}{2}\; \big [ \gamma_3 
      + |\gamma_1 - \gamma_2|\, \big ] \; . \]
For that purpose, we substitute $x= \cos\theta$ to obtain a quadratic 
expression:
\[ A\; x^2 \pm \lambda\, a_3\; x + \gamma_3 -A \ge 0 \; , \quad 
   A= \frac{\lambda^2}{2}\, \big [ \gamma_3 + |\gamma_1 - \gamma_2|\, \big ] 
 \; . \]
The LHS describes a parabola. Therefore, the inequality holds, if we can prove
that the equation $A\; x^2 \pm \lambda\, a_3\; x + \gamma_3 -A = 0$ has no
solution or at most one solution. For that purpose we consider the discriminant
and show that it is less or equal to zero. For later convenience, we define 
$g_\pm = \lambda_3 \pm |\gamma_1 - \gamma_2|$. Then we may write:
\begin{align} 
\lambda^2\, a_3^2 - 4A\, (\gamma_3 - A)  &\le 0 \notag\\
\Leftarrow\quad a_3^2 - g_+\, (2\gamma_3 - \lambda^2\, g_+) &\le 0 \; , \quad
A= \frac{\lambda^2\, g_+}{2}\notag\\
\Leftarrow\quad g_+\, g_- -2g_+\, \gamma_3 + \lambda^2 g_+^2 &\le 0\; , \quad
a_3^2 = g_+\, g_-\notag\\
\Leftarrow\quad g_- -2\gamma_3 + \lambda^2 g_+  &\le 0 \notag\\
\Leftarrow\quad -g_+ (1 - \lambda^2) &\le 0 \; , \quad g_- - 2\lambda_3 = -g_+
\; .
\notag\end{align}
This completes the proof. The discriminant is negative semidefinite. Therefore
the two inequalities in Eq.~(\ref{aEP:tau12cond}) are always fulfilled and
can be omitted.

\section{\label{aPP} Positivity of the non-unital anisotropic Pauli channel}

We start from the condition, given in Eq.~(\ref{EP:geomtauinequal}),
\[ \vec\rme_r\, \cdot\, (\, \underline\gamma\, \vec\rme_r + \vec\tau\, ) \ge 0
   \; , \]
where $\vec\rme_r$ is a unit vector in spherical coordinates, parametrized by
the angles $\theta,\beta$. We aim at constructing the surface $\mathcal{T}$ 
which forms the outer boundary of the region of points $\vec\tau$, where the 
above inequality holds (note that this region contains the origin 
$\vec\tau = \vec o$, and that it must be convex%
\footnote{For fixed $R$, two different quantum generators 
$\mathcal{L}_1, \mathcal{L}_2$ are given by $\vec\tau_1$ and $\vec\tau_2$, and 
any intermediate generator 
$\lambda\, \mathcal{L}_1 + (1-\lambda)\, \mathcal{L}_2$ is given by 
$\lambda\, \vec\tau_1 + (1-\lambda)\, \vec\tau_2$.}%
). The condition for $\vec\tau \in \mathcal{T}$ can be cast into the following
set of equations:
\begin{align}
 \vec\rme_r \cdot (\, \underline\gamma\, \vec\rme_r + \vec\tau\, ) &= 0 
\notag\\
 \frac{\partial}{\partial\theta}\; 
 \vec\rme_r \cdot (\, \underline\gamma\, \vec\rme_r + \vec\tau\, ) &= 0
\label{aPP:outsurf}\\
 \frac{\partial}{\partial\beta}\; 
 \vec\rme_r \cdot (\, \underline\gamma\, \vec\rme_r + \vec\tau\, ) &= 0\; .
\notag\end{align}
The argument is as follows: Consider the LHS of the first equation as a 
function $f(\vec\tau,\theta,\beta)$, then we may compute
\[ f_{\rm max}(\vec\tau) = \max_{\theta,\beta} f(\vec\tau,\theta,\beta) \; , \]
by finding the critical points (there may be more than one) 
$(\theta_i, \beta_i)$, where the last two equalities of Eq.~(\ref{aPP:outsurf}) 
hold.  Typically, for some fixed but arbitrary point $\vec\tau$, some of the 
values of $\{\, f(\vec\tau,\theta_i,\beta_i)\, \}$ may be positive and others 
negative; some may correspond to local maxima, others to local minima, and 
still others may correspond neither to one nor to the other group. However, the 
global maximum will always be among these points.

The calculation of the partial derivatives is simplified by the fact that
\begin{align}
 \frac{\partial\, \vec\rme_r}{\partial\theta} &= 
   \begin{pmatrix}
     \cos\theta\, \cos\beta\\ \cos\theta\, \sin\beta\\ -\sin\theta
   \end{pmatrix} = \vec\rme_\theta \; , \notag\\
   \frac{\partial\, \vec\rme_r}{\partial\beta} &= 
   \begin{pmatrix}
     -\sin\theta\, \sin\beta\\ \sin\theta\, \cos\beta\\ 0
   \end{pmatrix} = \sin\theta\, \vec\rme_\beta \; , 
\notag\end{align}
such that $\{ \vec\rme_r , \vec\rme_\theta , \vec\rme_\beta\, \}$ form a 
system of orthonormal vectors. Therefore the system of equations in 
Eq.~(\ref{aPP:outsurf}) becomes
\begin{align}
 \vec\rme_r \cdot (\, \underline\gamma\, \vec\rme_r + \vec\tau\, ) &= 0 
\notag\\
 \vec\rme_\theta \cdot (\, \underline\gamma\, \vec\rme_r + \vec\tau\, ) +
 \vec\rme_r \cdot \underline\gamma\, \vec\rme_\theta &= 0
\notag\\
 \vec\rme_\beta \cdot (\, \underline\gamma\, \vec\rme_r + \vec\tau\, ) +
 \vec\rme_r \cdot \underline\gamma\, \vec\rme_\beta &= 0\; , 
\notag\end{align}
which is equivalent to
\begin{align}
 \vec\rme_r \cdot (\, \underline\gamma\, \vec\rme_r + \vec\tau\, ) &= 0 
\notag\\
\vec\rme_\theta \cdot (\, 2\, \underline\gamma\, \vec\rme_r + \vec\tau\, ) &= 0 
\notag\\
\sin\theta\; \vec\rme_\beta \cdot (\, 2\, \underline\gamma\, \vec\rme_r 
 + \vec\tau\, ) &= 0 \; .
\notag\end{align}
We started by asking for which points $\vec\tau$, there exist a critical point
$(\theta_i,\beta_i)$ corresponding to a global maximum such that this set of 
equations is fulfilled. That point would then fore sure belong to the desired
surface $\mathcal{T}$. 
However, starting from this relation, we may say that it assigns to any pair
of angles $(\theta,\beta)$, a unique $\vec\tau$, such that that pair of angles
is a critical point (of any nature), while $f(\vec\tau,\theta,\beta) = 0$. That
means that for $\vec\tau\in\mathcal{T}$, it is a necessary but not sufficient
condition that it satisfies this equation for some pair of angles 
$(\theta,\beta)$. Therefore, the surface $\mathcal{T}$ must be a subset of the
set of solutions $\vec\tau$ to this equation.

The last to equalities imply that $2\underline{\gamma}\vec\rme_r +\vec\tau =
\alpha\, \vec\rme_r$ for some unknown real parameter $\alpha$. Inserting this
into the first equality, we obtain
\[ \vec\rme_r \cdot (\, \alpha\, \vec\rme_r - \underline{\gamma}\vec\rme_r\, ) 
   = 0 \quad\Rightarrow\quad 
   \alpha = \vec\rme_r \cdot \underline{\gamma}\vec\rme_r \; , \]
and finally
\begin{align}
\vec\tau = (\vec\rme_r \cdot \underline{\gamma}\, \vec\rme_r)\; \vec\rme_r 
   - 2\, \underline\gamma\, \vec\rme_r \; . 
\label{aPP:finres}\end{align}

\section{Canonical form of quantum process generators}
\label{sec:canonicalform}
In this appendix we prove that the dissipation matrix $D$, introduced in 
Eq.~(\ref{CP:LindMastEq}) is unitarily equivalent to $C_\perp$, defined in 
Sec.~\ref{CM}. To do this we first prove that any generator of a trace-preserving map, as defined in Eq.~\ref{CP:LocalInterMap}, can always be written as Eq.~\ref{CP:LindMastEq}, proving that the matrices $D$ and $H$ always exist. 

Notice that the generator $\mathcal{L}$ preserves hermiticity, thus it has an hermitian Choi matrix $C_\mathcal{L}$ of $d^2\times d^2$. One can trivially write such matrix as
\begin{equation}
C_\mathcal{L}=C_\phi-\proj{\Psi}{\Phi_\text{B}}-\proj{\Phi_\text{B}}{\Psi},
\label{eq:general_hermitian_form}
\end{equation}
where $\ket{\Psi}=-\omega_{\perp}C_\mathcal{L}\ket{\Phi_\text{B}}-\frac{\lambda}{2} 
\ket{\Phi_\text{B}}$, $\lambda=\bra{\Phi_\text{B}}C_\mathcal{L}\ket{\Phi_\text{B}}$ and 
$\omega_\perp C_\mathcal{L} \omega_\perp=\omega_{\perp}C_\phi\omega_{\perp}=C_\phi$~\cite{WoEiCuCi08}. To shorten the notation, we introduce the projector on the Bell state as
$\omega= |\Phi_{\rm B}\ra\la\Phi_{\rm B}|$, and the projector on the 
complementary subspace as $\omega_\perp = \One - \omega$.  Observe that choosing the operator basis $\lbrace F_i \rbrace_{1\leq i\leq d^2}$, as introduced in the main text, it is simple to prove that the matrix $C_\phi$ can be understood also as the Choi matrix of the following superoperator:
\begin{equation}
\phi\left[\rho\right]=\sum_{i,j=1}^{d^2-1} D_{i j} F_{i}\rho F^{\dagger}_{j},
\end{equation}
with $D$ hermitian. This can be shown observing that the summation $\sum_{i,j=1}^{d^2-1}$ goes over only trace-less operators, thus
\[ \omega \left(\left(\text{id} \otimes \phi\right)[\omega]\right)
   = \sum_{i,j=1}^{d^2-1}D_{ij}\frac{1}{d} \text{tr} 
     \left(F_{i}\right)\omega \left(\One\otimes F^{\dagger}_{j}\right)
   = \mathbf{0} \; , \]
and similarly 
\[ \left(\left(\text{id} \otimes \phi[\omega]\right)\right)\omega=\mathbf{0}
   \; .  \] 
For the second and third terms of Eq.~\ref{eq:general_hermitian_form} it is easy to show that the corresponding superoperator is simply 
$\rho \mapsto-\kappa \rho -\rho \kappa^\dagger$, thus we identify $\ket{\Psi}=d\left(\One\otimes \kappa \right)\ket\Phi_\text{B}$.

Up to now we have shown that hermiticity preserving generators have the form 
\begin{equation}
\mathcal{L}\left[\rho\right]=\phi\left[\rho\right]-\kappa\rho -\rho \kappa^{\dagger}.
\end{equation}
Using the trace-preserving condition of the quantum map we can write down explicitly the form of the hermitian part of $\kappa$. Such condition is translated to $\mathcal{L}$ as $\mathcal{L}^*[\One]=\mathbf{0}$, where ``$*$'' indicates the adjoint map of $\mathcal{L}$ under the Hilbert-Schmidt inner product. Thus we have
$$\kappa+\kappa^\dagger=\phi^*[\One],$$
\ie{} the hermitian part of $\kappa$ is given by $\frac{1}{2} \sum_{i,j=1}^{d^2-1} D_{i j} F_j^\dagger F_{i}$. Simply writing the antihermitian part as $i H$ we end up with
$$\kappa=\rmi H+ 
\frac{1}{2} \sum_{i,j=1}^{d^2-1} D_{i j} F_j^\dagger F_{i}.$$
Therefore any generator defined as in Eq.~\ref{CP:LocalInterMap} has the form depicted in Eq.~\ref{CP:LindMastEq}~\cite{WoEiCuCi08,Evans1977}, and we identify $C_\phi=C_\perp$.
Additionally notice that the the superoperator $\phi$ is CP iff $D\geq 0$~\cite{HeiZimBook11}, thus $D\geq 0 \Leftrightarrow C_\perp\geq 0$. In such case $\mathcal{L}$ has Lindblad form.

Now we prove that $D$ and $C_\perp$ are related by an unitary conjugation. Choosing an arbitrary basis orthogonal to $\ket{\Phi_\text{B}}$, say $\lbrace | \phi_i \rangle \rbrace_{i=1}^{d^2-1}$, the matrix $C_\perp$ in such basis has the entries 

\begin{align}
\left(C_\perp\right)_{nm}&=\bra{\phi_n} C_\perp \ket{\phi_m}\nonumber\\
&=d\langle \phi_n |\left(\text{id}\otimes \phi [\omega]\right)|\phi_m \rangle\nonumber\\
&=d\sum_{i,j=1}^{d^2-1}a_i a_j D_{ij} \langle \phi_n|\Phi^{(i)}\rangle \langle \Phi^{(j)}| \phi_m \rangle,
\label{eq_conjug_1} 
\end{align}
with $|\Phi^{(i)}\rangle=\frac{1}{a_i}\mathbf{1}\otimes F_i |\Phi_\text{B}\rangle$ and $a_i= | \text{id}\otimes F_i |\Phi_\text{B} \rangle|^2$.
Now observe that $a_i a_j\langle \Phi^{(i)}|\Phi^{(j)} \rangle=\bra{\Phi_\text{B}}\One\otimes F_i^\dagger F_j\ket{\Phi_\text{B}}$ is equal to $\frac{1}{d}\text{tr}\left(F_i^\dagger F_j\right)=\frac{1}{d}\delta_{ij}$, thus $a_i=\frac{1}{\sqrt{d}} \forall i=1\dots d^2$. Defining $V_{ni}=\langle\phi_n|\Phi^{(i)}\rangle$ and substituting the value of $a_i$ in Eq.~\ref{eq_conjug_1} we end up with
$$C_\perp=V D V^{\dagger},$$
with $V$ unitary, given that $\ket{\phi_n}$ and $\ket{\Phi^{(i)}}$ are properly normalized quantum states.

Choosing simply $\ket{\phi_i}=\ket{\Phi^{(i)}}$, we have simply that $C_\perp=D$. 

Recalling that knowing the matrix $D$, we can compute directly the hermitian part of $\kappa$ by calculating $\phi^{*}[\One]$. To compute $H$ we can get $\kappa$ using $\ket{\Psi}$, and subtract its hermiant part. In particular using the canonical basis, we have that $\langle mn | \Psi \rangle=d \bra{mn} \One \otimes \kappa \ket{\Psi_\text{B}}=\sqrt{d} \bra{n}k\ket{m}= \sqrt{d}\kappa_{nm}$. Once getting $\kappa$, we have that $H=- \rmi \kappa + \frac{1}{2} \rmi \phi^{*}[\One]$.

\end{appendix}

%\section*{References}

%\bibliography{/home/thomas/Documents/Bib/JabRef-Deli}
\bibliography{/home/gorin/Documentos/Bib/JabRef-Deli,DavidRefs}

\end{document}